\documentclass{emulateapj}
\usepackage{natbib}
\newcommand{\tx}[1]{\textrm{#1}}

\newcommand{\kms}{km~$\tx{s}^{-1}$}

\newcommand{\hi}{\ion{H}{1}}
\newcommand{\vlsr}{\ensuremath{V_{LSR}}}
\newcommand{\gbtidl}{{\sc GBTIDL}}
\newcommand{\vegas}{{\sc VEGAS}}

\slugcomment{ApJL, in prep.}
\shorttitle{Antlia~B}
\shortauthors{Sand et al.}

\begin{document}
 \title{Antlia~B: A faint dwarf galaxy member of the NGC~3109 association}

\author{D. J. Sand,$\!$\altaffilmark{1} K. Spekkens,$\!$\altaffilmark{2} D. Crnojevi\'{c},$\!$\altaffilmark{1} J.~R. Hargis,$\!$\altaffilmark{3} B. Willman,$\!$\altaffilmark{3} J. Strader,$\!$\altaffilmark{3} C.J. Grillmair$\!$\altaffilmark{5}} \email{david.sand@ttu.edu}

\begin{abstract}

We report the discovery of Antlia~B, a faint dwarf galaxy at a projected distance of $\sim$72 kpc from NGC~3109 ($M_{V}$$\sim$$-$15 mag), the primary galaxy of the NGC~3109 dwarf association at the edge of the Local Group.  The tip of the red giant branch distance to Antlia~B is $D$=1.29$\pm$0.10 Mpc, which is consistent with the distance to NGC~3109.  A qualitative analysis indicates the new dwarf's stellar population has both an old, metal poor red giant branch ($\gtrsim$10 Gyr, [Fe/H]$\sim$$-$2), and a younger blue population with an age of $\sim$200-400 Myr, analogous to the original Antlia dwarf, another likely satellite of NGC~3109.  Antlia~B has \ion{H}{1} gas at a velocity of $v_{helio,HI}$=376 km s$^{-1}$, confirming the association with NGC~3109 ($v_{helio}$=403 km s$^{-1}$).  The \ion{H}{1} gas mass (M$_{HI}$=2.8$\pm$0.2$\times$10$^{5}$ M$_{\odot}$), stellar luminosity ($M_{V}$=$-$9.7$\pm$0.6 mag) and half light radius ($r_{h}$=273$\pm$29 pc) are all consistent with the properties of dwarf irregular and dwarf spheroidal galaxies in the Local Volume, and is most similar to the Leo~P dwarf galaxy.  The discovery of Antlia~B is the initial result from a Dark Energy Camera survey for halo substructure and faint dwarf companions to NGC~3109 with the goal of comparing observed substructure with expectations from the $\Lambda$+Cold Dark Matter model in the sub-Milky Way regime.

\end{abstract}
 
\altaffiltext{1}{Texas Tech University, Physics Department, Box 41051, Lubbock, TX 79409-1051, USA}
\altaffiltext{2}{Royal Military College of Canada, Department of Physics, PO Box 17000, Station Forces, Kingston, Ontario, Canada K7K 7B4}
\altaffiltext{3}{Department of Astronomy, Haverford College, 370 Lancaster Avenue, Haverford, PA 19041, USA }

\altaffiltext{4}{Michigan State University, Department of Physics and Astronomy, East Lansing, MI 48824, USA}
\altaffiltext{5}{Spitzer Science Center, 1200 E. California Blvd., Pasadena, CA 91125, USA}

\section{Introduction}

The faint end of the galaxy luminosity function is important for understanding the astrophysics of the $\Lambda$+Cold Dark Matter ($\Lambda$CDM) picture of galaxy formation.  Over the last $\sim$15 years, observational work has uncovered a population of very faint dwarf galaxies around the Milky Way \citep[MW; e.g.][and references therein]{Willman10} and M31 \citep[e.g.~][among others]{Martin13}, while many more should be discovered by upcoming surveys \citep[e.g.][]{Hargis14}.  At the same time numerical simulations indicate that there may be hundreds more subhalos that are effectively `dark'.
There are many physical mechanisms that are expected to either keep small dark matter halos dark, or severely limit the number of stars they form \citep[e.g.~ultraviolet heating from reionization, supernova feedback, tidal/ram pressure stripping; see e.g.][and many others]{Maccio10,Penarrubia12,Pontzen12,Bovy12,Arraki12}, but it is difficult to disentangle these effects by only looking at the dwarf galaxy population in a single environment -- the Local Group.

The search for faint satellites around the MW continues to flourish, most recently due to the Dark Energy Camera (DECam) at Cerro Tololo International Observatory (CTIO) and the accompanying Dark Energy Survey (DES), as well as other DECam wide-field surveys \citep[although non-DECam surveys have also been successful, e.g., Pan-STARRS;][]{Laevens15,Laevens15b}.   Since the beginning of 2015, $\sim$10 new MW dwarf satellites have been found with DECam \citep{Bechtol15,Koposov15,Kim15,Martin15}.  Of particular interest has been the possible association of these new dwarf systems with the Large Magellanic Cloud \citep{Koposov15,Deason15}.  Faint dwarfs associated with larger dwarf galaxies are expected in the $\Lambda$CDM model \citep[e.g.][]{Donghia08,Wheeler15}. However, it will be difficult to definitively associate a given satellite galaxy with the Magellanic Cloud system as opposed to the MW.  Searching for faint dwarf galaxy systems around nearby, isolated dwarf galaxy groups is a more promising route to understanding the role that parent galaxy mass plays in determining faint satellite properties.




Motivated to push near-field cosmology beyond the Local Group, we have completed a wide-field survey around NGC~3109 \citep[$M_{V}$$\sim$$-$15, $D$$\sim$1.28 Mpc;][]{McConnachie12}, the most prominent galaxy in the `NGC~3109 dwarf association' \citep{Tully06}.  Other members of the dwarf association include the Antlia dwarf, Sextans A and B, and possibly the recently discovered Leo~P \citep[e.g.][]{Giovanelli13}.  Undertaken with DECam at CTIO, our goals were to find further faint satellite companions to NGC~3109 and to study the structure and metallicity of its halo.   The complete dataset consists of 15 DECam pointings (45 deg$^2$), giving partial coverage out to $R$$\sim$100 kpc from NGC~3109's center. Typical image depths are $r$$\sim$24.5 mag, which is $\sim$2 magnitudes below the tip of the red giant branch (TRGB) at $D$$\sim$1.3 Mpc.

During inspection of the incoming data at the telescope, we spotted a new dwarf galaxy companion to NGC~3109, which we have dubbed Antlia~B (Figure~\ref{fig:dwarf}), corresponding to its constellation and the prior existence of the Antlia dwarf.  Here we present this new discovery.  Future work will present a resolved stellar analysis of the halo of NGC~3109, as well as {\it Hubble Space Telescope} (HST) imaging of Antlia~B and other faint dwarf galaxy candidates (PID: 14078; PI: J. Hargis).  In \S~\ref{sec:datareduce} we present our observational dataset, while in \S~\ref{sec:properties} we present our physical measurements of Antlia~B.  We discuss Antlia~B in context and conclude in \S~\ref{sec:conclude}.





\section{Observations} \label{sec:datareduce}

\subsection{Optical Broadband Imaging}

The broadband data presented in this paper were taken on 2015 March 13 (UT) with DECam \citep{DECam} on the  Blanco 4m telescope (2015A-0130; PI: D. Crnojevi\'{c}).  
DECam has 62 2K$\times$4K CCDs arranged in a hexagonally shaped mosaic.  The pixel scale of 0\farcs27 per pixel yields a total field of view of 3 square degrees.  All of the imaging data were reduced using the DECam community pipeline \citep{Valdes14}, and we directly use the stacked, world coordinate system-corrected $g$ and $r$ band images.  The final image stacks consisted of 7$\times$300 s and 7$\times$150 s images in the $g$ and $r$ bands, respectively, with image point spread functions of $\sim$0\farcs9 for both.  Small dithers between individual exposures were used to cover the chip gaps in the final stacks.

We extracted a single `tile' containing Antlia~B ($\sim$45'$\times$30' in size) for our analysis.  Stellar photometry was performed using a methodology similar to previous work on Local Group/Volume dwarfs \citep[e.g.][]{Sand12,Sand14} with the {\sc DAOPHOTII/Allstar} package \citep{Stetson94}.

Several SDSS fields at varying airmasses were imaged every night of our NGC~3109 campaign, and we utilized the Radiometric All-Sky Infrared Camera \citep[][]{RASICAM} data in the image headers to verify that the sky was clear at each position.  Calibration zeropoints, linear color terms and extinction coefficients were derived from the SDSS fields in order to convert instrumental magnitudes to the SDSS photometric system. Although the entire NGC~3109 DECam dataset was not photometric, the relevant Antlia~B pointing was, and so we applied our derived zeropoints and color terms directly to our stellar catalogs. 

Once the photometry was calibrated, we performed artificial star tests (utilizing the {\sc DAOPHOT} routine {\sc ADDSTAR}) to determine our photometric errors and completeness.
Over many iterations, we injected $\sim$5$\times$$10^{5}$ artificial stars into our data, with an $r$-band magnitude range of 18 to 30 and a $g-r$ color of $-$0.5 to 2.0, and re-analyzed these data in an identical fashion as the unaltered data.  The 50\% (90\%) completeness level was at $r$=24.7 (23.6) and $g$=25.3 (24.5) mag.  The data do not suffer from crowding incompleteness until $r$$\sim$24.5 in the very central regions of Antlia~B.  For the purpose of this work, we simply derive our structural parameters for Antlia~B with a magnitude cutoff of $r$=24.5 mag (along with consistency checks) so as to avoid any issues (see \S~\ref{sec:struct}).

The final photometric catalog was corrected for Galactic extinction \citep{Schlafly11} star by star, with a typical color excess of E(B-V)$\approx$0.069 mag.  All magnitudes presented in this work have this correction applied.  In Figure~\ref{fig:dwarf} we show the CMD of Antlia~B within its derived half-light radius (\S~\ref{sec:struct}), along with several equal-area `background' CMDs.  We discuss the stellar populations in Antlia~B in \S~\ref{sec:pop}.

\subsection{$H\alpha$ Imaging}

Narrowband H$\alpha$ imaging of Antlia~B was obtained on 2015 June 12 (UT) with the Goodman High-Throughput Spectrograph \citep[][]{Goodman} on the SOAR telescope.  The observations (taken in 1\farcs5--2\farcs0 seeing conditions) included 3$\times$300 s in the H$\alpha$ filter ($\lambda$$_{central}$=6565 \AA, FWHM = 65 \AA) and 4$\times$120 s in the $r$ band for continuum subtraction.   Standard image processing and calibration steps were carried out to produce a continuum-subtracted H$\alpha$ image and flux limits \citep[][]{Kennicutt08}.  No H$\alpha$ emission was detected in association with Antlia~B, to a 3$\sigma$ point source detection limit of $\sim$2.1$\times$10$^{-16}$ erg s$^{-1}$cm$^{-2}$.  

\subsection{Green Bank Telescope Observations}\label{gbtinfo}

We obtained director's discretionary time on the Robert C. Byrd Green Bank Telescope (GBT) in 2015 June (Program AGBT15A\_384; PI: K. Spekkens) to carry out position-switched \hi\ observations of Antlia~B. We used the Versatile GBT Astronomical Spectrometer (\vegas) with a bandpass of $11.72\,$MHz and $0.4\,$kHz channels centered at $\vlsr=500\,$\kms~in the \hi\ line to obtain spectra along the line of sight to Antlia~B as well as at several reference locations. The total integration time of 35 mins was divided equally between the on- and off-target locations, the latter serving to flatten the spectral baseline in the target spectrum at the processing stage. The 9.1 arcmin full-width at half-maximum beam of the GBT at this observing frequency well exceeds the half-light stellar diameter of Antlia~B ($\sim$1.4', see \S~\ref{sec:pop}), and we therefore expect any \hi\ in this system to be recovered in a single GBT pointing. 

The data were reduced using the standard \gbtidl\footnote{ \tt http://gbtidl.nrao.edu/} routine {\it getps} and smoothed to a spectral resolution of $2\,$\kms. The resulting spectrum is shown in Fig.~\ref{fig:HIspec}: we detect \hi\ emission along the line of sight to Antlia~B with S/N$\sim$15. The profile peak at $v_{helio}$$\sim$375 km s$^{-1}$ indicates that Antlia~B has a very similar velocity to NGC~3109 ($v_{helio,HI}$=403 km s$^{-1}$).

\section{Properties of Antlia~B}\label{sec:properties}

\subsection{Distance}\label{sec:distance}

We measure the distance to Antlia~B with the TRGB method \citep[e.g.][]{Lee93,Rizzi07,Crnojevic14}.  A Sobel edge detection filter is used to identify the sharp transition in the $r$-band luminosity function that coincides with the brightest, metal poor RGB stars which can be used as a standard candle.  We find $r_{0,TRGB}$ = 22.55$\pm$0.06 mag for Antlia~B, which corresponds to a distance modulus of $(m-M_{0})$ = 25.56$\pm$0.16 mag ($D$=1.29$\pm$0.10 Mpc), after adopting a TRGB absolute value of $M^{TRGB}_{r}$=$-$3.01$\pm$0.1 \citep[][]{Sand14}.  This value is identical to the distance to NGC~3109 \citep[$D$=1.28$\pm$0.02 Mpc;][]{Dalcanton09}, to within the uncertainties.  Considering both the compatible distance and HI velocity between Antlia~B and NGC~3109, we conclude that they are associated.

\subsection{Stellar Population}\label{sec:pop}

Inspection of the CMD of Antlia~B reveals a complex star formation history (SFH).  In Figure~\ref{fig:dwarf}, we show several representative theoretical isochrones (\citet{Dotter08} for the 13.5 Gyr stellar populations, and \citet{Girardi10} for the younger isochrones) over-plotted onto a CMD from within the half light radius, along with several background CMDs drawn from equal areas in outlying regions.  The data are consistent with having an old, metal poor stellar population ($>$10 Gyr, [Fe/H]$\sim$$-$2), with a younger, more metal-rich component ($\sim$200--400 Myr, [Fe/H]$\sim$$-$1).  The younger stellar populations are coincident with relatively heavy foreground star contamination, but the group of stars at $g-r$$\sim$$-$0.5 and $r$$\sim$23.5 are consistent with a $\sim$100-200 Myr stellar population, with no counterpart in the background CMDs.  In the CMD region occupied by the $\sim$200-400 Myr isochrones, the number of stars in the Antlia~B CMD outnumber those seen in the background CMDs by a factor of $\sim$2. There is no indication of a younger stellar population ($\lesssim$100 Myr), which concurs with our H$\alpha$ non-detection.  No GALEX imaging is available at the position of Antlia~B.

Antlia~B's SFH is analogous to that of the original Antlia dwarf \citep{McQuinn10}, which displays an old, metal poor stellar population ($\sim$10 Gyr), as well as a moderately young component of $\sim$100--400 Myr.  There is no very young stellar population in Antlia ($<$10 Myr) in deep HST photometry \citep{McQuinn10}, and no H$\alpha$ detection \citep{Lee09}.  Several other so-called `transition' systems (e.g.~DDO~210 and LGS~3), have analogous star formation histories as Antlia and Antlia~B, with no very recent star formation despite the presence of \ion{H}{1} gas \citep[e.g.][]{Weisz11}.  Upcoming {\it HST} observations will provide a clearer picture of the SFH of Antlia~B, which will allow for more direct comparisons between the two systems.

\subsection{Stellar Structure \& Luminosity}\label{sec:struct}

\begin{figure*}
\begin{center}
\mbox{ \epsfysize=7.0cm \epsfbox{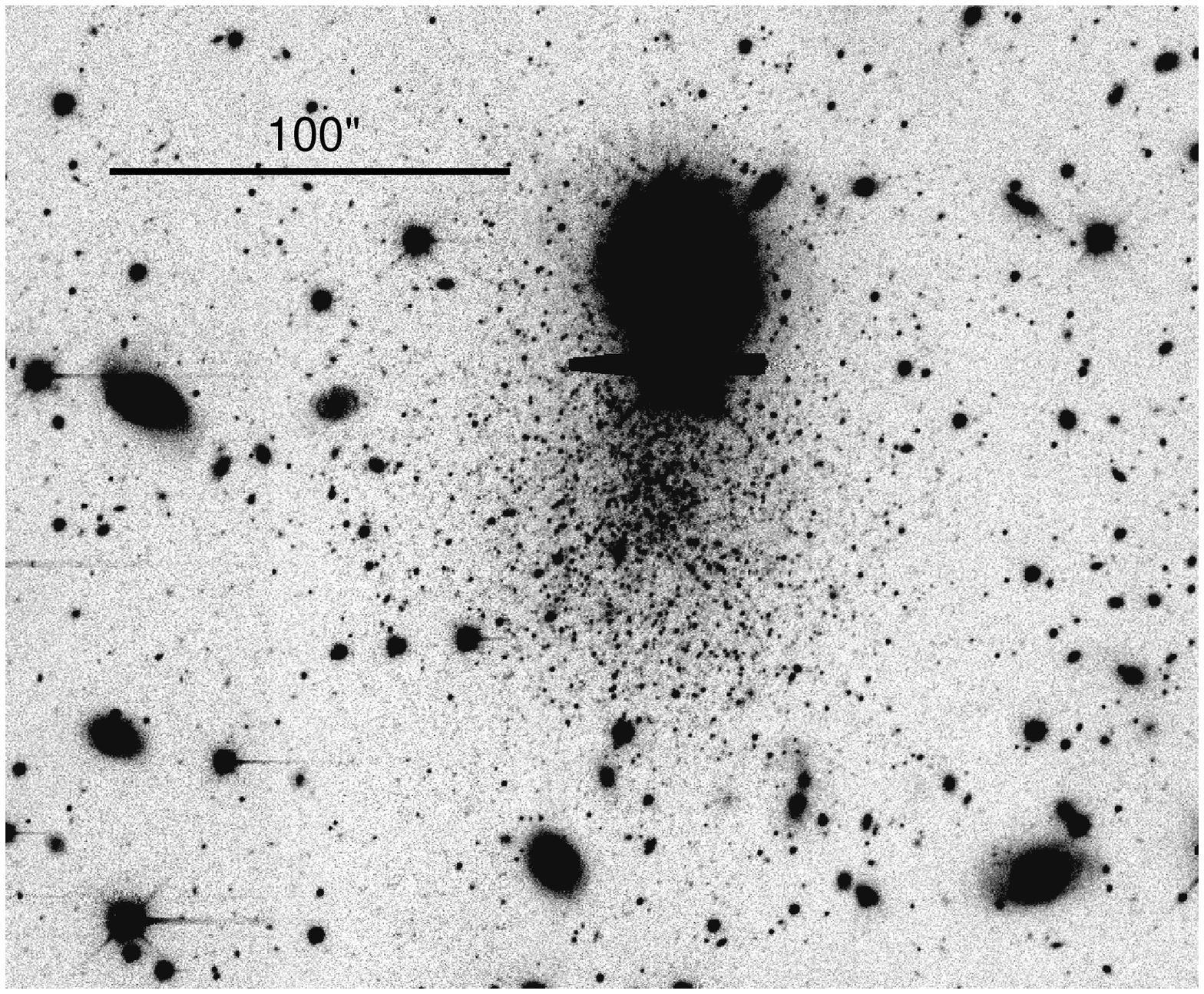}} 
\mbox{ \epsfysize=9.0cm \epsfbox{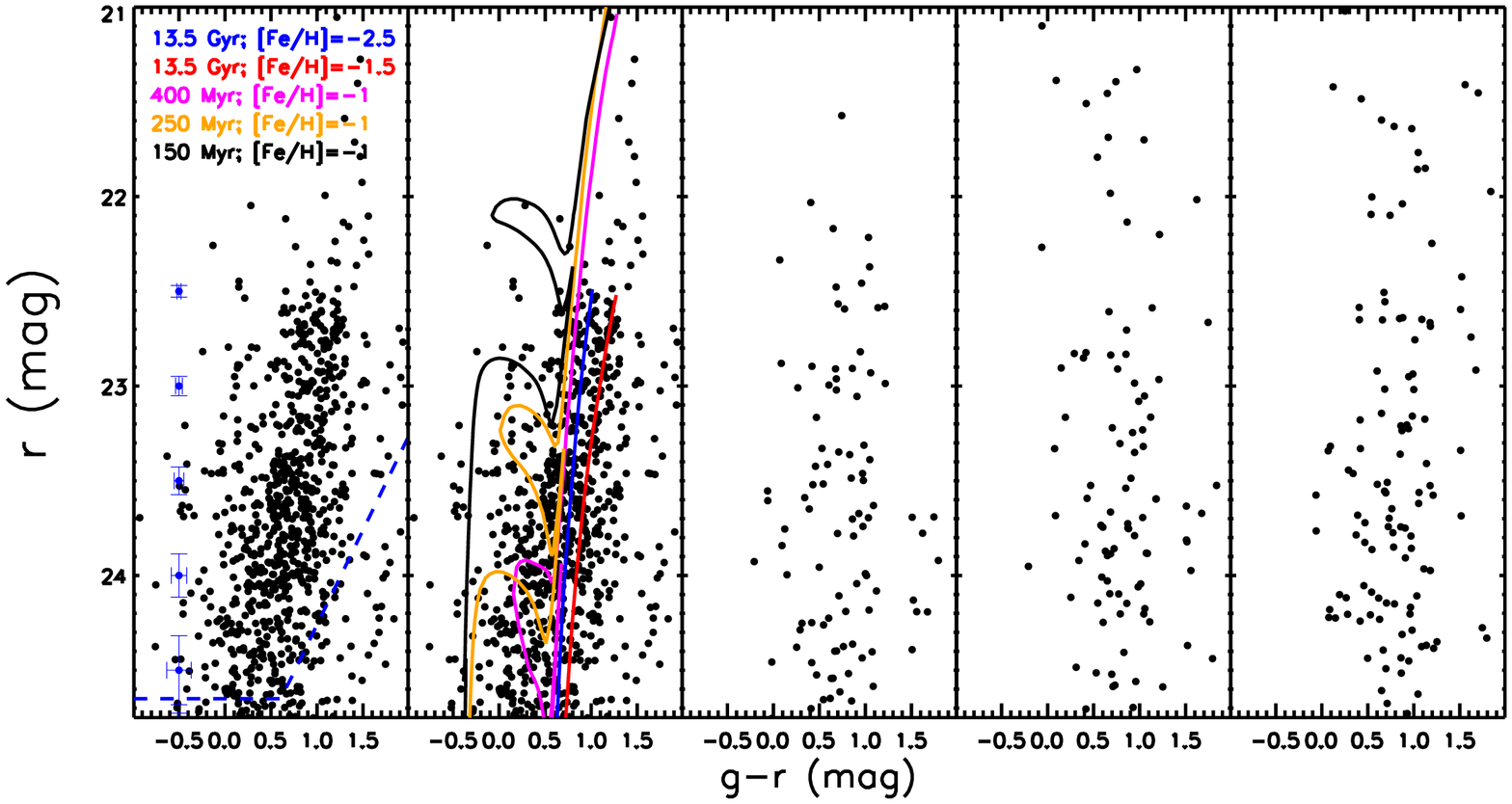}} 
\caption{  {\bf Top:} DECam $r$-band image of Antlia~B.  
Note the saturated foreground star and background spiral galaxy which partially overlaps Antlia~B in the Northern direction. North is up and East is to the left.  {\bf Bottom:} In the two left panels, we display our CMD of Antlia~B within 1 $r_{h}$ (43\farcs2).  Along the left side of the far left CMD are the typical uncertainties at different $r$-band magnitudes, as determined via artificial star tests.  The blue dashed line shows the 50\% completeness limit.   In the second panel we plot several representative theoretical isochrones, indicating that Antlia~B has both an old, metal poor stellar population ($>$10 Gyr, [Fe/H]$\sim$$-$2) and a younger, more metal-rich component ($\sim$200--400 Myr, [Fe/H]$\sim$$-$1). See \S~\ref{sec:pop} for  a discussion.  The three right panels show CMDs from random equal-area regions, illustrating typical ``background" CMDs. \label{fig:dwarf}}
\end{center}
\end{figure*}

\begin{figure*}
\begin{center}
\mbox{ \epsfysize=7.0cm \epsfbox{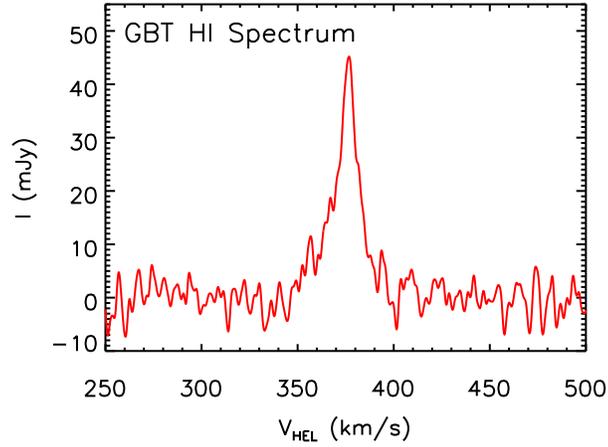}} 
\caption{   Position-switched, $2\,$\kms-resolution \hi\ spectrum obtained with the GBT along the line of sight to Antlia~B. The profile's spectral location corresponds to $v_{helio,HI}$=376$\pm$2 km s$^{-1}$ and the implied \hi\ properties are consistent with expectations for a gas-rich dIrr satellite of NGC~3109 (which is at $v_{helio}$=403 km s$^{-1}$).  
\label{fig:HIspec}}
\end{center}
\end{figure*}

We determine the structural parameters of Antlia~B with the same maximum likelihood technique utilized for Local Group dwarfs 
\citep[e.g.][]{sdssstruct}.  Stars which were consistent with old red giant branch isochrones placed at the distance of Antlia~B were included in the structural analysis (but not explicitly stars from the younger `blue loop' populations) down to $r$=24.5 mag. We fit an exponential profile plus a constant background to the data with the central position ($\alpha_{0}$,$\delta_{0}$), position angle ($\theta$), ellipticity ($\epsilon$), half-light radius ($r_{h}$) and background surface density as free parameters.  Uncertainties on each parameter were calculated via bootstrap resampling the data 1000 times.  Careful masking of both the saturated star and background spiral galaxy to the North of Antlia~B (Figure~\ref{fig:dwarf}) ensured that these regions were de-weighted in the maximum likelihood calculation.  As a check on our results, we repeated the calculations while only including RGB stars down to $r$=23.5, and found that the derived structural parameters did not change to within the uncertainties; we thus only report the parameters measured from our `deep' maximum likelihood run.   The results of the structural analysis are presented in Table~\ref{table:properties}.  Antlia~B has an exponential half-light radius of $r_{h}$=273$\pm$29 pc, and an ellipticity of $\epsilon$=0.30$\pm$0.05, both of which are comparable to the MW's classical dwarf spheroidals.

We estimate Antlia~B's luminosity directly via aperture photometry \citep[see][]{Sand14,Sand15}.  We use an elliptical aperture matched to the half light radius, ellipticity and position angle of Antlia~B and randomly place 100 equal area apertures throughout the DECam tile to estimate the background.  For the aperture centered on Antlia~B itself, we only take flux from the Southern half of the ellipse and multiply it by a factor of two, assuming symmetry.  This avoids the contamination from the foreground saturated star and background spiral galaxy (Figure~\ref{fig:dwarf}).  Accounting for the fact that our aperture also only covers half the total flux of Antlia~B, we find $M_{g}$ = $-$9.4$\pm$0.4 and $M_{r}$ = $-$9.9$\pm$0.4, where the uncertainty was calculated based on the scatter in measurements from the background apertures, as well as the 
uncertainty in the distance modulus.  Using the filter transformations of \citet{Jordi06}, this leads to $M_{V}$=$-$9.7$\pm$0.6 mag.


\subsection{HI Content}

We measure the systemic velocity $V_{helio,HI}$, velocity width at 50\% of the profile peak $W50_{HI}$, and the \hi\ line flux $S_{21}$ for Antlia~B from the \hi\ spectrum in Fig.~2 following the approach detailed in \citet{Springob05}, and tabulate the resulting values in Table~1. Adopting D=1.29$\pm$0.1 Mpc, the $S_{21}$ measurement corresponds to an \hi\ mass of $M_{HI} = (2.8 \pm 0.2) \times 10^5\,\mathrm{M_\odot}$, where the uncertainty includes contributions from the \hi\ spectrum and TRGB distance added in quadrature. The \hi\ mass to stellar luminosity ratio of Antlia~B is therefore $M_{HI}/L_V \sim 0.4 (M_\odot/L_\odot)$, implying that Antlia~B is a gas-rich dwarf irregular galaxy. We compare Antlia~B's \hi\ gas fraction to other Local Volume dwarf irregulars in \S~4. 

\section{Discussion and Conclusions}\label{sec:conclude}

We have presented the discovery of Antlia~B, a faint dwarf galaxy $R$$\sim$72 kpc in projection from the center of NGC~3109 ($D$=1.28 Mpc; $v_{\odot}$=403 km s$^{-1}$).  Antlia~B's inferred distance (1.29$\pm$0.1 Mpc) and \ion{H}{1} velocity ($v_{\odot}$=376 km s$^{-1}$) make the physical association with NGC~3109 certain.  The stellar population of Antlia~B is complex, with an old and metal poor component ($\gtrsim$10 Gyr; [Fe/H]$\sim$$-$2) as well as a younger stellar population ($\sim$200-400 Myr).   We detect $M_{HI}$ = (2.8$\pm$0.2)$\times$10$^5$ $M_{\odot}$ of HI gas in Antlia~B, implying that it is a gas-rich dwarf irregular galaxy.

\begin{figure*}
\begin{center}
\mbox{ \epsfysize=6.1cm \epsfbox{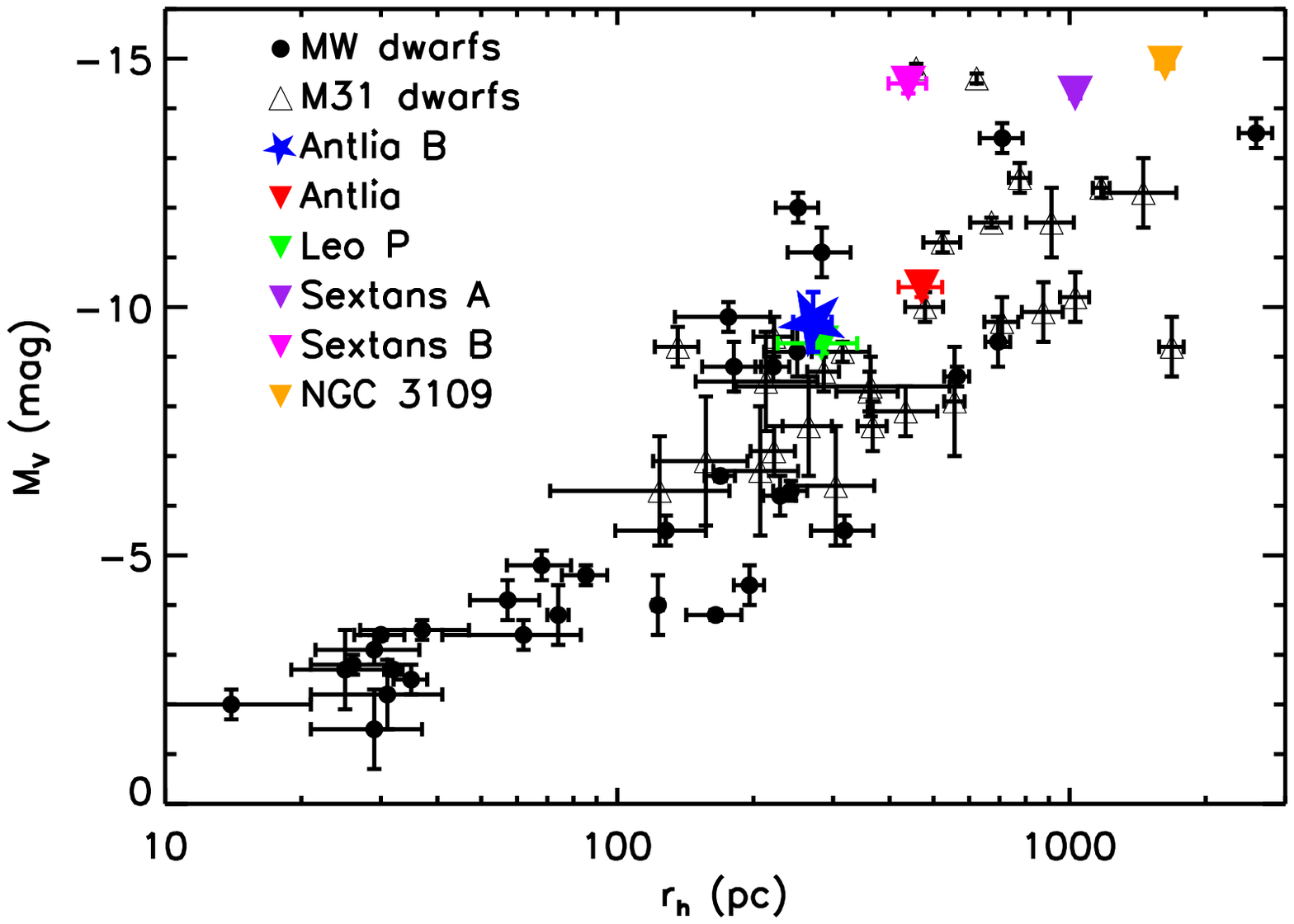}} 
\mbox{ \epsfysize=6.3cm \epsfbox{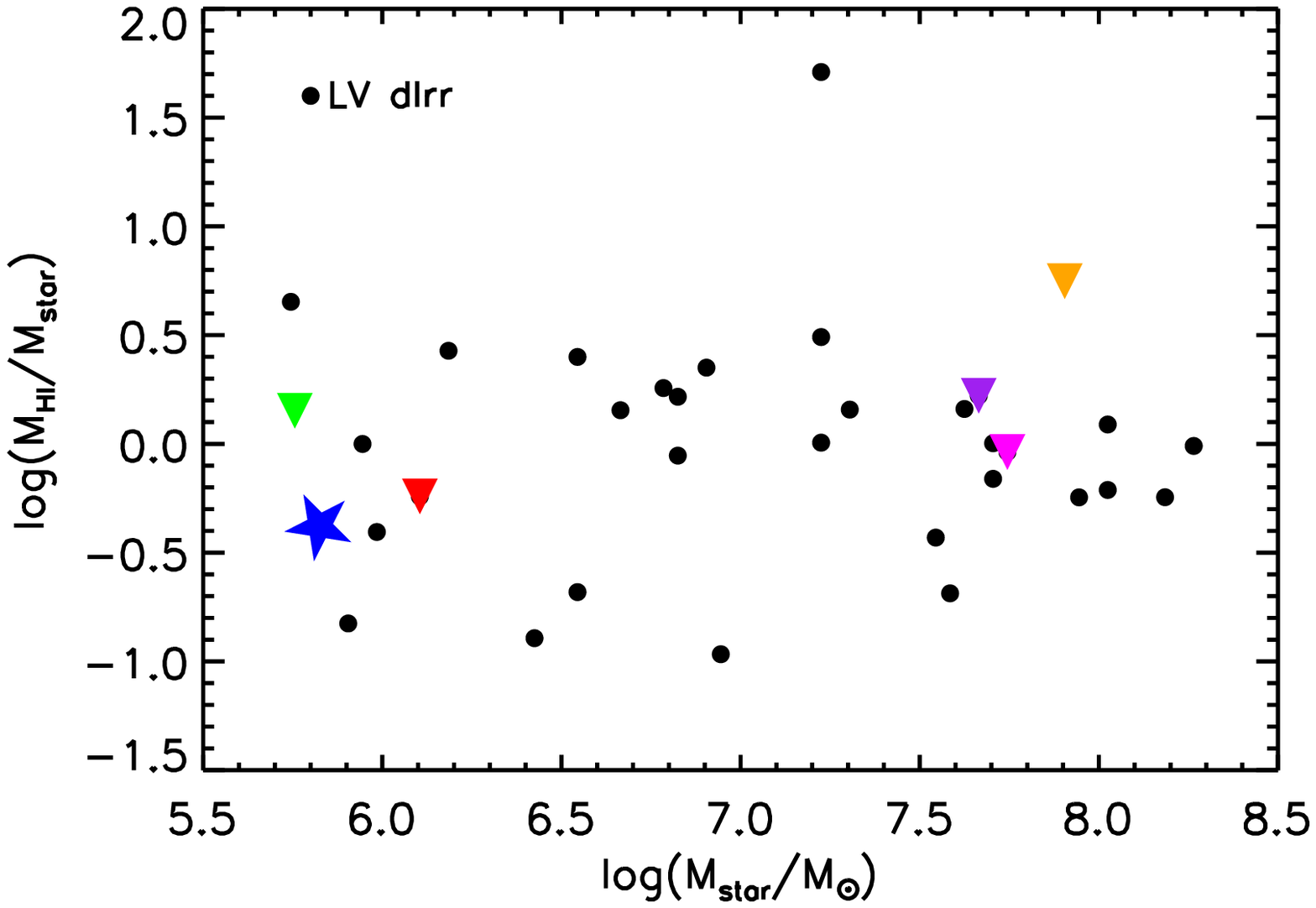}} 
\caption{  {\bf Left:} Absolute magnitude as a function of half light radius of Antlia~B and the rest of the NGC~3109 association.  Also plotted are Local Group dwarf galaxies \citep{McConnachie12,Sand12,Koposov15}.  Antlia~B is similar to Leo~P, while the NGC~3109 association as a whole is consistent with the dwarfs in the Local Group. {\bf Right:} The ratio of $M_{HI}$/$M_{star}$ as a function of stellar mass for the NGC~3109 association and Local Volume dwarf irregulars.  Antlia~B has a typical $M_{HI}$/$M_{star}$ for gas rich, low stellar mass members of the Local Volume.  \label{fig:props}}
\end{center}
\end{figure*}

We compare the physical properties of Antlia~B with other members of the NGC~3109 association and Local Volume in Figure~\ref{fig:props}.  To compute the \ion{H}{1} gas mass to stellar mass ratio (which probes the efficiency of gas conversion to stars) we have assumed that ($M_{star}$/$L_{V}$)/($M_{\odot}$/$L_{\odot}$)=1 and translated our absolute $V$-band magnitudes.  Antlia~B's properties are broadly consistent with other Local Volume members of its size and luminosity.  In particular, it is most similar to Leo~P \citep[][]{McQuinn13}, albeit with a slightly lower HI gas mass fraction.  

Given their proximity to NGC~3109, both Antlia (projected separation of $\sim40\,$kpc) and Antlia~B (projected separation of $\sim70\,$kpc) may be surprisingly gas-rich.  Galaxies with similar stellar masses to NGC~3109 ($M_{star} \sim 1 \times 10^8\,M_\odot$) in the ELVIS simulation suite (Garrison-Kimmel et al. 2014) have dark matter halo virial radii of $\sim100\,$kpc.  Therefore, unless the line-of-sight separations of Antlia and Antlia~B are at least as large as their projected ones (and keeping in mind that both Antlia and Antlia~B's line of sight distance uncertainty is $\sim$100 kpc), they lie within the virial radius of their parent galaxy. Satellites within the MW's virial radius with similar stellar masses to Antlia and Antlia~B are devoid of \hi\ \citep{Spekkens14}, consistent with a scenario in which ram pressure stripping from the MW's hot gas corona has swept away their interstellar media during past pericentric passages \citep[e.g.][]{Gatto13}, although other physical mechanisms may contribute to this gas removal \citep[e.g.][for recent discussions]{Grcevich09,Spekkens14}. The \hi\ richness of Antlia and Antlia~B therefore suggests that they have not yet reached a point in their orbits where ram pressure stripping is effective: this could be because they have not yet had a pericentric passage, or because either their orbital speeds are not high enough or NGC~3109's hot corona isn't dense enough at pericenter for the mechanism to have much effect. 

Recently it has been suggested that the NGC~3109 dwarf association displays a linear configuration in both spatial and velocity coordinates, suggestive of an infalling filament or tidal origin \citep{Bellazzini13,Pawlowski14}.  While Antlia~B does appear to fall on this proposed linear structure, this is necessarily the case because of the geometry of our survey, which extended out to $R$$\sim$100 kpc from NGC~3109.

The substructure properties of sub-MW mass halos is an unobserved portion of parameter space, but simulations suggest that they will have a wealth of substructure \citep[e.g.][]{Wheeler15}.  The discovery of Antlia~B, and upcoming work on NGC~3109's overall substructure properties, will provide tests of the $\Lambda$+CDM picture of galaxy formation in a new mass regime.

\acknowledgments
  DJS  acknowledges support from
NSF grant AST-1412504.  KS acknowledges support from the Natural Sciences and Engineering Research Council of Canada.  B.W. and J.H. were supported by an NSF Faculty Early Career
Development (CAREER) award (AST-1151462).  The National Radio Astronomy Observatory is a facility of the National Science Foundation operated under cooperative agreement by Associated Universities, Inc.  This work was supported in part by National Science Foundation Grant No. PHYS-1066293 and the Aspen Center for Physics.  This project used data obtained with the Dark Energy Camera, which was constructed by the Dark Energy Survey collaboration. Funding for the DES Projects has been provided by the DOE and NSF(USA), MISE(Spain), STFC(UK), HEFCE(UK). NCSA(UIUC), KICP(U. Chicago), CCAPP(Ohio State), MIFPA(Texas A\&M), CNPQ, FAPERJ, FINEP (Brazil), MINECO(Spain), DFG(Germany) and the collaborating institutions in the DES, which are Argonne Lab, UC Santa Cruz, University of Cambridge, CIEMAT-Madrid, University of Chicago, University College London, DES-Brazil Consortium, University of Edinburgh, ETH Zurich, Fermilab, University of Illinois, ICE (IEEC-CSIC), IFAE Barcelona, Lawrence Berkeley Lab, LMU Munchen and the associated Excellence Cluster Universe, University of Michigan, NOAO, University of Nottingham, Ohio State University, University of Pennsylvania, University of Portsmouth, SLAC National Lab, Stanford University, University of Sussex, and Texas A\&M University.

\bibliographystyle{apj}

\clearpage

\begin{deluxetable*}{lcccccccccc}
\tablecolumns{2}
\tablecaption{Antlia~B Properties\label{table:properties}}
\tablehead{
\colhead{Parameter}  & \colhead{Value} \\
}\\
\startdata
RA$_{0}$ (h:m:s) & 09:48:56.08 $\pm$2.1"\\
DEC$_{0}$ (d:m:s) & -25:59:24.0 $\pm$3.8"\\
$m-M$ (mag) & 25.56$\pm$0.16\\
D (Mpc) &  1.29$\pm$0.1\\
$M_{V}$ (mag) & $-$9.7$\pm$0.6 \\
$r_{h}$ (arcsec) & 43.2$\pm$4.2 \\
$r_{h}$ (pc) & 273$\pm$29 \\
$\epsilon$ & 0.30$\pm$0.05\\
$\theta$ (deg) & 4.0$\pm$12.0 \\
$S_{21}$ (Jy km s$^{-1}$) & 0.72$\pm$0.05\\
$W50_{HI}$ (km s$^{-1}$) & 17$\pm$ 4\\
$M_{HI}$ (10$^5$ $M_{\odot}$) & 2.8 $\pm$ 0.2\\
$v_{helio, HI}$ (km s$^{-1}$)& 376 $\pm$ 2 
\enddata
\label{tab:props}
\end{deluxetable*}


\end{document}